\documentclass[12pt]{article}
\usepackage{amssymb,amsfonts}
\newlength{\extraspace}
\newlength{\extraspaces}
\newcommand{\be}{\begin{equation}
\addtolength{\abovedisplayskip}{\extraspaces}
\addtolength{\belowdisplayskip}{\extraspaces}
\addtolength{\abovedisplayshortskip}{\extraspace}
\addtolength{\belowdisplayshortskip}{\extraspace}}
\newcommand{\ee}{\end{equation}}

\newcommand{\ba}{\begin{eqnarray}
\addtolength{\abovedisplayskip}{\extraspaces}
\addtolength{\belowdisplayskip}{\extraspaces}
\addtolength{\abovedisplayshortskip}{\extraspace}
\addtolength{\belowdisplayshortskip}{\extraspace}}
\newcommand{\ea}{\end{eqnarray}}

\newcommand{\newsection}[1]{
\vspace{15mm}
\pagebreak[3]
\addtocounter{section}{1}
\setcounter{equation}{0}
\setcounter{subsection}{0}
\setcounter{footnote}{0}
\begin{flushleft}
{\large\bf \thesection. #1}
\end{flushleft}
\nopagebreak
\medskip
\nopagebreak}

\newcommand{\Tr}{{\rm Tr}}

\begin{document}

\addtolength{\baselineskip}{.8mm}

{\thispagestyle{empty}

\noindent \hspace{1cm}  \hfill Revised version \hspace{1cm}\\
\mbox{}                 \hfill November 2004 \hspace{1cm}\\

\begin{center}
\vspace*{1.0cm}
{\large\bf On the loop--loop scattering amplitudes} \\
{\large\bf in Abelian and non--Abelian gauge theories} \\
\vspace*{1.0cm}
{\large Enrico Meggiolaro\footnote{E--mail:
enrico.meggiolaro@df.unipi.it} }\\
\vspace*{0.5cm}{\normalsize
{Dipartimento di Fisica,
Universit\`a di Pisa, \\
Largo Pontecorvo 3,
I--56127 Pisa, Italy.}}\\
\vspace*{2cm}{\large \bf Abstract}
\end{center}

\noindent
The high--energy elastic scattering amplitude of two colour--singlet
$q \bar{q}$ pairs is governed by the correlation function of two Wilson loops,
which follow the classical straight lines for quark (antiquark) trajectories.
This quantity is expected to be free of IR divergences, differently from what
happens for the parton--parton elastic scattering amplitude, described, in the
high--energy limit, by the expectation value of two Wilson lines. We shall
explicitly test this IR finitness by a direct non--perturbative computation of
the loop--loop scattering amplitudes in the (pedagogic, but surely physically
interesting) case of {\it quenched} QED. The results obtained for the Abelian
case will be generalized to the case of a non--Abelian gauge theory with $N_c$
colours, but stopping to the order ${\cal O}(g^4)$ in perturbation theory.
In connection with the above--mentioned IR finitness, we shall also discuss
some analytic properties of the loop--loop scattering amplitudes in both
Abelian and non--Abelian gauge theories, when going from Minkowskian to
Euclidean theory, which can be relevant to the still unsolved problem of the
$s$--dependence of hadron--hadron total cross--sections.

\vspace{0.5cm}
\noindent
(PACS codes: 11.15.-q, 11.55.Bq, 11.80.Fv, 12.38.-t, 12.38.Aw, 12.38.Bx,
13.85.-t)
}
\vfill\eject

\newsection{Introduction}

\noindent
The high--energy parton--parton elastic scattering amplitude can be described,
in the center--of--mass reference system, by the expectation value of two
{\it infinite} Wilson lines, running along the classical trajectories of the
colliding particles \cite{Nachtmann91,Nachtmann97,Meggiolaro96,Meggiolaro01}.
However, the above description suffers from infrared (IR) divergences, which
are typical of $3 + 1$ dimensional gauge theories. As was first pointed out by
Verlinde and Verlinde in \cite{Verlinde}, the IR singularity is originated by
the fact that the trajectories of the Wilson lines were taken to be lightlike
and therefore have an infinite distance in rapidity space.
One can regularize this IR problem by giving the Wilson lines a small
timelike component, such that they coincide with the classical trajectories
for quarks with a non--zero mass $m$ (this is equivalent to consider two Wilson
lines forming a certain {\it finite} hyperbolic angle $\chi$ in Minkowskian
space--time; of course, $\chi \to \infty$ when $s \to \infty$),
and, in addition, by considering {\it finite} Wilson lines, extending in
proper time from $-T$ to $T$ (and eventually letting $T \to +\infty$)
\cite{Verlinde,Meggiolaro02}.

Such a regularization of the IR singularities gives rise to a non--trivial
$s$--dependence of the amplitudes, as obtained by ordinary perturbation theory
\cite{Cheng-Wu-book,Lipatov} and as confirmed by the experiments on
hadron--hadron scattering processes (see, for example, Ref. \cite{pomeron-book}
and references therein). We refer the reader to Refs.
\cite{Verlinde} and \cite{Meggiolaro97,Meggiolaro98} for a detailed discussion
about this point.
However, the parton--parton scattering amplitudes, while being
finite at any given value of $T$, are divergent in the limit $T \to \infty$.
In some cases this IR divergence can be factorized out and one thus ends up
with an IR--finite (physical) quantity: this is what happens, for example,
in the case of {\it quenched} QED \cite{Meggiolaro96,Meggiolaro97}.

A way to get rid of the problem of the IR--cutoff dependence is to consider
an IR--finite physical quantity, like the elastic scattering amplitude of two
colourless states in gauge theories, e.g., two $q \bar{q}$ meson states.
It was shown in Refs. \cite{Nachtmann97,Dosch,Berger} that the high--energy
meson--meson elastic scattering amplitude can be approximately reconstructed
by first evaluating, in the eikonal approximation, the elastic scattering
amplitude of two $q \bar{q}$ pairs (usually called ``{\it dipoles}''), of
given transverse sizes $\vec{R}_{1\perp}$ and $\vec{R}_{2\perp}$ respectively,
and then averaging this amplitude over all possible dipole transverse
separations $\vec{R}_{1\perp}$ and $\vec{R}_{2\perp}$ with two proper squared
wave functions $|\psi_1 (\vec{R}_{1\perp})|^2$ and
$|\psi_2 (\vec{R}_{2\perp})|^2$, describing the two interacting mesons.
It turns out that the high--energy elastic scattering amplitude of two
$q \bar{q}$ pairs of transverse sizes $\vec{R}_{1\perp}$ and
$\vec{R}_{2\perp}$ is governed by the correlation function of two Wilson loops
${\cal W}_1$ and ${\cal W}_2$, which follow the classical straight lines for
quark (antiquark) trajectories (and are closed by straight--line paths at
proper times $\tau = \pm T$, with $T \to \infty$):
\be
{\cal T}_{(ll)} (s,t;~\vec{R}_{1\perp},\vec{R}_{2\perp}) \equiv -i~2s
\displaystyle\int d^2 \vec{z}_\perp e^{i \vec{q}_\perp \cdot \vec{z}_\perp}
\left[ {\langle {\cal W}_1 {\cal W}_2 \rangle \over
\langle {\cal W}_1 \rangle \langle {\cal W}_2 \rangle} -1 \right] ,
\label{scatt-loop}
\ee
where $s$ and $t = -\vec{q}_\perp^2$ ($\vec{q}_\perp$ being the tranferred
momentum) are the usual Mandelstam variables.
The expectation values $\langle f(A) \rangle$ in Eq. (\ref{scatt-loop})
are averages of $f(A)$ in the sense of the functional integration over the
gluon field $A^\mu$ (including also the determinant of the fermion matrix).
More explicitly the Wilson loops ${\cal W}_1$ and ${\cal W}_2$ are so defined:
\ba
{\cal W}^{(T)}_1 (\vec{z}_\perp,\vec{R}_{1\perp}) &\equiv&
{1 \over N_c} \Tr \left\{ {\cal P} \exp
\left[ -ig \displaystyle\oint_{{\cal C}_1} A_\mu(x) dx^\mu \right] \right\} ,
\nonumber \\
{\cal W}^{(T)}_2 (\vec{0}_\perp,\vec{R}_{2\perp}) &\equiv&
{1 \over N_c} \Tr \left\{ {\cal P} \exp
\left[ -ig \displaystyle\oint_{{\cal C}_2} A_\mu(x) dx^\mu \right] \right\} ,
\label{QCDloops}
\ea
where $A_\mu = A_\mu^a T^a$, $T^a$ ($a = 1,\ldots,N_c^2-1$ in a non--Abelian
gauge theory with $N_c$ colours) being the generators of the $SU(N_c)$ algebra
in the fundamental representation (with the standard normalization
$\Tr [T^a T^b] = \delta_{ab}/2$).
Moreover, ${\cal P}$ denotes the ``{\it path ordering}'' along the given path
${\cal C}$; ${\cal C}_1$ and ${\cal C}_2$ are two rectangular paths which
follow the classical straight lines for quark [$X_{(+)}(\tau)$, forward in
proper time $\tau$] and antiquark [$X_{(-)}(\tau)$, backward in $\tau$]
trajectories, i.e.,
\ba
{\cal C}_1 &\to&
X_{(\pm 1)}^\mu(\tau) = z^\mu + {p_1^\mu \over m} \tau
\pm {R_1^\mu \over 2} , \nonumber \\
{\cal C}_2 &\to&
X_{(\pm 2)}^\mu(\tau) = {p_2^\mu \over m} \tau \pm {R_2^\mu \over 2} ,
\label{traj}
\ea
and are closed by straight--line paths at proper times $\tau = \pm T$.
Here $p_1$ and $p_2$ are the four--momenta of the two quarks and of the two
antiquarks with mass $m$, moving with speed $\beta$ and $-\beta$ along, for
example, the $x^1$--direction:
\ba
p_1 &=& E (1,\beta,0,0) = m (\cosh {\chi \over 2},\sinh {\chi \over 2},0,0) ,
\nonumber \\
p_2 &=& E (1,-\beta,0,0) = m (\cosh {\chi \over 2},-\sinh {\chi \over 2},0,0) ,
\label{p1p2}
\ea
where $\chi = 2~{\rm arctanh} \beta$ is the hyperbolic angle [in the
{\it longitudinal} plane $(x^0,x^1)$ in Minkowski space--time] between the two
trajectories $X^\mu_{(+1)}(\tau)$ and $X^\mu_{(+2)}(\tau)$.
Moreover, $R_1 = (0,0,\vec{R}_{1\perp})$, $R_2 = (0,0,\vec{R}_{2\perp})$,
where $\vec{R}_{1\perp}$ and $\vec{R}_{2\perp}$ are the {\it transverse} sizes
[i.e., separations in the {\it transverse} plane $(x^2,x^3)$] of the two loops
and $z = (0,0,\vec{z}_\perp)$, where $\vec{z}_\perp = (z^2,z^3)$ is the
impact--parameter distance between the two loops in the {\it transverse} plane.

As we have said above, in the case of the loop--loop scattering amplitudes
the dependence on the IR cutoff $T$ is expected to be removed (for $T \to
\infty$) together with the related IR divergence which was present for the
case of Wilson lines.
As a pedagogic example to illustrate these considerations (but, of course,
also interesting by itself for physical reasons!), in Sect. 2 we shall consider
the simple case of {\it quenched} QED. In this case, the calculation of the
(normalized) correlator of the two Wilson loops $\langle {\cal W}_1 {\cal W}_2
\rangle / \langle {\cal W}_1 \rangle \langle {\cal W}_2 \rangle$ can be
performed explicitly both in Minkowskian and in Euclidean theory
(taking, in this last case, two IR--regularized Euclidean Wilson loops forming
a certain angle $\theta$ in Euclidean four--space) and one finds that the
two quantities are indeed IR finite (when $T \to \infty$) and are
connected by the same analytic continuation in the angular variables
($\chi \to i\theta$) and in the IR cutoff ($T \to -iT$) which was already
derived in the case of Wilson {\it lines}
\cite{Meggiolaro02,Meggiolaro97,Meggiolaro98}.
In Sect. 3 we shall generalize the results obtained in Sect. 2 to the case of
a non--Abelian gauge theory with $N_c$ colours, but stopping to the order
${\cal O}(g^4)$ in perturbation theory.
Finally, in Sect. 4 we shall furnish a sketch of the
proof (based on the detailed proof reported in Ref. \cite{Meggiolaro02} for
the case of Wilson lines) that the above--mentioned relation of analytic
continuation, between the loop--loop scattering amplitudes in the Minkowskian
and in the Euclidean theory, is indeed an {\it exact} and {\it general} result,
i.e., not restricted to some order in perturbation theory or to some other
approximation (such as the {\it quenched} approximation), valid both
for the Abelian and the non--Abelian case.\\
We shall conclude the paper with some general considerations about the IR
finitness of the loop--loop scattering amplitudes (at the end of Sect. 4)
and the use of the Euclidean--to--Minkowskian analytic continuation to address
the (still unsolved) problem of the asymptotic $s$--dependence of
hadron--hadron total cross sections (Sect. 5).

\newsection{The Abelian case}

\noindent
In this section we shall discuss the Abelian case.
The electromagnetic Wilson loops are defined as in the non--Abelian case,
after replacing $g$ with $e$, the electric coupling--constant (electric
charge), and the gluon field with the photon field $A^\mu$:
\ba
{\cal W}^{(T)}_1 (\vec{z}_\perp,\vec{R}_{1\perp}) &=&
\exp \left[ -ie \displaystyle\oint_{{\cal C}_1} A_\mu(x) dx^\mu \right],
\nonumber \\
{\cal W}^{(T)}_2 (\vec{0}_\perp,\vec{R}_{2\perp}) &=&
\exp \left[ -ie \displaystyle\oint_{{\cal C}_2} A_\mu(x) dx^\mu \right].
\label{loops-def}
\ea
Thanks to the simple form of the Abelian theory (in particular to the 
absence of self--interactions among the photons), it turns out that 
it is possible to explicitly evaluate the expectation values
$\langle {\cal W}^{(T)}_1 \rangle$, $\langle {\cal W}^{(T)}_2 \rangle$ and
$\langle {\cal W}^{(T)}_1 {\cal W}^{(T)}_2 \rangle$, at least in the
so--called {\it quenched} approximation, where vacuum polarization effects, 
arising from the presence of loops of dynamical fermions, are neglected.
This amounts to setting the fermionic determinant to a constant in the
functional integrals, which are then reduced to {\it Gaussian} functional
integrals and can be evaluated by well--known simple techniques (see, e.g.,
Refs.  \cite{Meggiolaro96} and \cite{Meggiolaro97}).
One thus obtains the following result:
\be
{ \langle {\cal W}^{(T)}_1 (\vec{z}_\perp,\vec{R}_{1\perp})
{\cal W}^{(T)}_2 (\vec{0}_\perp,\vec{R}_{2\perp}) \rangle \over
\langle {\cal W}^{(T)}_1 (\vec{z}_\perp,\vec{R}_{1\perp}) \rangle
\langle {\cal W}^{(T)}_2 (\vec{0}_\perp,\vec{R}_{2\perp}) \rangle }
= \exp \left[ i (\Phi_1^{(T)} + \Phi_2^{(T)}) \right] ,
\label{corr12}
\ee
where the term $\Phi_1^{(T)}$ in the exponent is equal to
\ba
\lefteqn{
\Phi_1^{(T)} = 16e^2 \cosh\chi \displaystyle\int {d^4 k \over (2\pi)^4}
{e^{-i\vec{k}_\perp \cdot \vec{z}_\perp} \over k^2 +i\varepsilon}
\sin \left( {\vec{k}_\perp \cdot \vec{R}_{1\perp} \over 2} \right)
\sin \left( {\vec{k}_\perp \cdot \vec{R}_{2\perp} \over 2} \right) }
\nonumber \\
& & \times {\sin[(k^0 \cosh{\chi \over 2} - k^1 \sinh{\chi \over 2})T] \over
(k^0 \cosh{\chi \over 2} - k^1 \sinh{\chi \over 2})}
{\sin[(k^0 \cosh{\chi \over 2} + k^1 \sinh{\chi \over 2})T] \over
(k^0 \cosh{\chi \over 2} + k^1 \sinh{\chi \over 2})} ,
\label{exp1T}
\ea
and has the following {\it finite} non--zero limit when letting the IR
cutoff $T$ going to infinity:
\be
\displaystyle\lim_{T \to \infty} \Phi_1^{(T)} = -4e^2 \coth\chi
\displaystyle\int {d^2 \vec{k}_\perp \over (2\pi)^2}
{e^{-i\vec{k}_\perp \cdot \vec{z}_\perp} \over \vec{k}_\perp^2}
\sin \left( {\vec{k}_\perp \cdot \vec{R}_{1\perp} \over 2} \right)
\sin \left( {\vec{k}_\perp \cdot \vec{R}_{2\perp} \over 2} \right) .
\label{exp1}
\ee
On the contrary, the second term $\Phi_2^{(T)}$ in the exponent at the
right--hand side of Eq. (\ref{corr12}) is equal to
\ba
\lefteqn{
\Phi_2^{(T)}
= -16e^2~ \vec{R}_{1\perp} \cdot \vec{R}_{2\perp} \displaystyle\int
{d^4 k \over (2\pi)^4}
{e^{-i\vec{k}_\perp \cdot \vec{z}_\perp} \over k^2 +i\varepsilon}
\sin \left(  {\vec{k}_\perp \cdot \vec{R}_{1\perp} \over 2} \right)
\sin \left( {\vec{k}_\perp \cdot \vec{R}_{2\perp} \over 2} \right) }
\nonumber \\
& & \times {\sin[(k^0 \cosh{\chi \over 2} - k^1 \sinh{\chi \over 2})T]
\over \vec{k}_\perp \cdot \vec{R}_{1\perp}}
{\sin[(k^0 \cosh{\chi \over 2} + k^1 \sinh{\chi \over 2})T]
\over \vec{k}_\perp \cdot \vec{R}_{2\perp}} ,
\label{exp2T}
\ea
and, when letting the IR cutoff $T$ going to infinity, it has trivially a limit
equal to zero. Therefore, we obtain the following {\it finite} result for the
expression written in Eq. (\ref{corr12}), in the limit when the IR cutoff $T$
goes to infinity:
\ba
\lefteqn{
\displaystyle\lim_{T \to \infty}
{ \langle {\cal W}^{(T)}_1 (\vec{z}_\perp,\vec{R}_{1\perp})
{\cal W}^{(T)}_2 (\vec{0}_\perp,\vec{R}_{2\perp}) \rangle \over
\langle {\cal W}^{(T)}_1 (\vec{z}_\perp,\vec{R}_{1\perp}) \rangle
\langle {\cal W}^{(T)}_2 (\vec{0}_\perp,\vec{R}_{2\perp}) \rangle } }
\nonumber \\
& & = \exp \left[ -i 4e^2 \coth \chi
\displaystyle\int {d^2 \vec{k}_\perp \over (2\pi)^2}
{e^{-i\vec{k}_\perp \cdot \vec{z}_\perp} \over \vec{k}_\perp^2}
\sin \left( {\vec{k}_\perp \cdot \vec{R}_{1\perp} \over 2} \right)
\sin \left( {\vec{k}_\perp \cdot \vec{R}_{2\perp} \over 2} \right) \right] .
\label{loops12}
\ea
We can now repeat step by step the above procedure in the Euclidean theory.
Let us consider the correlation function of two Euclidean Wilson loops
\ba
\tilde{\cal W}^{(T)}_1 (\vec{z}_\perp,\vec{R}_{1\perp}) &=&
\exp \left[ -ie \displaystyle\oint_{\tilde{\cal C}_1}
A^{(E)}_\mu(x_E) dx_{E\mu} \right] ,
\nonumber \\
\tilde{\cal W}^{(T)}_2 (\vec{0}_\perp,\vec{R}_{2\perp}) &=&
\exp \left[ -ie \displaystyle\oint_{\tilde{\cal C}_2}
A^{(E)}_\mu(x_E) dx_{E\mu} \right] ,
\label{loops-defE}
\ea
where $\tilde{\cal C}_1$ and $\tilde{\cal C}_2$ are two rectangular paths
which follow the following straight--line trajectories [forward in $\tau$ for
$X^{(+)}_E(\tau)$, backward in $\tau$ for $X^{(-)}_E(\tau)$]:
\ba
\tilde{\cal C}_1 &\to&
X^{(\pm 1)}_{E\mu}(\tau) = z_{E\mu} + {p_{1E\mu} \over m}
\tau \pm {R_{1E\mu} \over 2} , \nonumber \\
\tilde{\cal C}_2 &\to&
X^{(\pm 2)}_{E\mu}(\tau) = {p_{2E\mu} \over m} \tau
\pm {R_{2E\mu} \over 2} ,
\label{trajE}
\ea
and are closed by straight--line paths at proper times $\tau = \pm T$.
Using the notation $V = (V_1,V_2,V_3,V_4)$ for a Euclidean four--vector, then
$R_{1E} = (0,\vec{R}_{1\perp},0)$ and $R_{2E} = (0,\vec{R}_{2\perp},0)$,
where $\vec{R}_{1\perp}$ and $\vec{R}_{2\perp}$ are
the transverse sizes of the two loops, and $z_E = (0,\vec{z}_\perp,0)$,
where $\vec{z}_\perp$ is the impact--parameter distance
between the two loops. Moreover, in the Euclidean theory we {\it choose}
a reference frame in which the spatial vectors $\vec{p}_{1E}$ and 
$\vec{p}_{2E} = -\vec{p}_{1E}$ are along the $x_{E1}$ direction and,
moreover, $p_{1E}^2 = p_{2E}^2 = m^2$; that is:
\ba
p_{1E} &=& m (\sin{\theta \over 2}, 0, 0, \cos{\theta \over 2} ) , \nonumber \\
p_{2E} &=& m (-\sin{\theta \over 2}, 0, 0, \cos{\theta \over 2} ) ,
\label{p1p2E}
\ea
where $\theta$ is the angle formed by the two trajectories in the
$(x_{E1},x_{E4})$ plane in Euclidean four--space.
By virtue of the $O(4)$ symmetry of the Euclidean theory,
the value of $\theta$ can be taken in the range $[0,\pi]$.\\
Proceeding analogously to the Minkowskian case, one finds (in the
{\it quenched} approximation) the following result:
\be
{ \langle \tilde{\cal W}^{(T)}_1 (\vec{z}_\perp,\vec{R}_{1\perp})
\tilde{\cal W}^{(T)}_2 (\vec{0}_\perp,\vec{R}_{2\perp}) \rangle_E \over
\langle \tilde{\cal W}^{(T)}_1 (\vec{z}_\perp,\vec{R}_{1\perp}) \rangle_E
\langle \tilde{\cal W}^{(T)}_2 (\vec{0}_\perp,\vec{R}_{2\perp}) \rangle_E }
= \exp \left[ - (\tilde{\Phi}_1^{(T)} + \tilde{\Phi}_2^{(T)}) \right] ,
\label{corr12E}
\ee
where the term $\tilde{\Phi}_1^{(T)}$ in the exponent is equal to
\ba
\lefteqn{
\tilde{\Phi}_1^{(T)}
= 16e^2 \cos\theta \displaystyle\int {d^4 k_E \over (2\pi)^4}
{e^{-i\vec{k}_\perp \cdot \vec{z}_\perp} \over k_E^2}
\sin \left( {\vec{k}_\perp \cdot \vec{R}_{1\perp} \over 2} \right)
\sin \left( {\vec{k}_\perp \cdot \vec{R}_{2\perp} \over 2} \right) }
\nonumber \\
& & \times {\sin[(k_{E1}\sin{\theta \over 2} + k_{E4}\cos{\theta \over 2})T]
\over (k_{E1}\sin{\theta \over 2} + k_{E4}\cos{\theta \over 2})}
{\sin[(-k_{E1}\sin{\theta \over 2} + k_{E4}\cos{\theta \over 2})T] \over
(-k_{E1}\sin{\theta \over 2} + k_{E4}\cos{\theta \over 2})} ,
\label{exp1TE}
\ea
and has the following {\it finite} non--zero limit when letting the IR
cutoff $T$ going to infinity:
\be
\displaystyle\lim_{T \to \infty} \tilde{\Phi}_1^{(T)}
= 4e^2 \cot\theta \displaystyle\int {d^2 \vec{k}_\perp \over (2\pi)^2}
{e^{-i\vec{k}_\perp \cdot \vec{z}_\perp} \over \vec{k}_\perp^2}
\sin \left( {\vec{k}_\perp \cdot \vec{R}_{1\perp} \over 2} \right)
\sin \left( {\vec{k}_\perp \cdot \vec{R}_{2\perp} \over 2} \right) .
\label{exp1E}
\ee
On the contrary, the second term $\tilde{\Phi}_2^{(T)}$ in the exponent at the
right--hand side of Eq. (\ref{corr12E}) is equal to
\ba
\lefteqn{
\tilde{\Phi}_2^{(T)}
= 16e^2~ \vec{R}_{1\perp} \cdot \vec{R}_{2\perp} \displaystyle\int
{d^4 k_E \over (2\pi)^4} {e^{-i\vec{k}_\perp \cdot \vec{z}_\perp} \over k_E^2}
\sin \left( {\vec{k}_\perp \cdot \vec{R}_{1\perp} \over 2} \right)
\sin \left( {\vec{k}_\perp \cdot \vec{R}_{2\perp} \over 2} \right) }
\nonumber \\
& & \times {\sin[(k_{E1}\sin{\theta \over 2} + k_{E4}\cos{\theta \over 2})T]
\over \vec{k}_\perp \cdot \vec{R}_{1\perp}}
{\sin[(-k_{E1}\sin{\theta \over 2} + k_{E4}\cos{\theta \over 2})T]
\over \vec{k}_\perp \cdot \vec{R}_{2\perp}} ,
\label{exp2TE}
\ea
and, when letting the IR cutoff $T$ going to infinity, it has trivially a limit
equal to zero. Therefore, we obtain the following {\it finite} result for the
expression written in Eq. (\ref{corr12E}), in the limit when the IR cutoff $T$
goes to infinity:
\ba
\lefteqn{
\displaystyle\lim_{T \to \infty}
{ \langle \tilde{\cal W}^{(T)}_1 (\vec{z}_\perp,\vec{R}_{1\perp})
\tilde{\cal W}^{(T)}_2 (\vec{0}_\perp,\vec{R}_{2\perp}) \rangle_E \over
\langle \tilde{\cal W}^{(T)}_1 (\vec{z}_\perp,\vec{R}_{1\perp}) \rangle_E
\langle \tilde{\cal W}^{(T)}_2 (\vec{0}_\perp,\vec{R}_{2\perp}) \rangle_E } }
\nonumber \\
& & = \exp \left[ -4e^2 \cot \theta
\displaystyle\int {d^2 \vec{k}_\perp \over (2\pi)^2}
{e^{-i\vec{k}_\perp \cdot \vec{z}_\perp} \over \vec{k}_\perp^2}
\sin \left( {\vec{k}_\perp \cdot \vec{R}_{1\perp} \over 2} \right)
\sin \left( {\vec{k}_\perp \cdot \vec{R}_{2\perp} \over 2} \right) \right] .
\label{loops12E}
\ea
We stress the fact that both the Minkowskian quantity (\ref{loops12}) and
the Euclidean quantity (\ref{loops12E}) are IR {\it finite} when
$T \to \infty$, differently from the corresponding quantities
constructed with Wilson {\it lines}, which were evaluated in Refs.
\cite{Meggiolaro96,Meggiolaro97}.

Moreover, considering the ratios (\ref{corr12}) and (\ref{corr12E})
in the presence of a {\it finite} IR cutoff $T$, which from now on will be
denoted as
\ba
{\cal G}_M(\chi;~T;~\vec{z}_\perp,\vec{R}_{1\perp},\vec{R}_{2\perp}) &\equiv&
{ \langle {\cal W}^{(T)}_1 (\vec{z}_\perp,\vec{R}_{1\perp})
{\cal W}^{(T)}_2 (\vec{0}_\perp,\vec{R}_{2\perp}) \rangle \over
\langle {\cal W}^{(T)}_1 (\vec{z}_\perp,\vec{R}_{1\perp}) \rangle
\langle {\cal W}^{(T)}_2 (\vec{0}_\perp,\vec{R}_{2\perp}) \rangle } ,
\nonumber \\
{\cal G}_E(\theta;~T;~\vec{z}_\perp,\vec{R}_{1\perp},\vec{R}_{2\perp}) &\equiv&
{ \langle \tilde{\cal W}^{(T)}_1 (\vec{z}_\perp,\vec{R}_{1\perp})
\tilde{\cal W}^{(T)}_2 (\vec{0}_\perp,\vec{R}_{2\perp}) \rangle_E \over
\langle \tilde{\cal W}^{(T)}_1 (\vec{z}_\perp,\vec{R}_{1\perp}) \rangle_E
\langle \tilde{\cal W}^{(T)}_2 (\vec{0}_\perp,\vec{R}_{2\perp}) \rangle_E } ,
\label{GM-GE}
\ea
we find, after comparing the Minkowskian expressions (\ref{corr12}),
(\ref{exp1T}) and (\ref{exp2T}) with the corresponding
Euclidean expressions (\ref{corr12E}), (\ref{exp1TE}) and
(\ref{exp2TE}), that they are connected by the following analytic continuation
in the angular variables and in the IR cutoff:
\ba
{\cal G}_E(\theta;~T;~\vec{z}_\perp,\vec{R}_{1\perp},\vec{R}_{2\perp}) &=&
{\cal G}_M(\chi \to i\theta;~T \to -iT;
~\vec{z}_\perp,\vec{R}_{1\perp},\vec{R}_{2\perp}) ,
\nonumber \\
{\rm i.e.,}~~
{\cal G}_M(\chi;~T;~\vec{z}_\perp,\vec{R}_{1\perp},\vec{R}_{2\perp}) &=&
{\cal G}_E(\theta \to -i\chi;~T \to iT;
~\vec{z}_\perp,\vec{R}_{1\perp},\vec{R}_{2\perp}) .
\label{analytic}
\ea
This is exactly the same analytic continuation which was already proved to be
valid in the case of Wilson {\it lines}
\cite{Meggiolaro02,Meggiolaro97,Meggiolaro98}.
It is also important to notice that the two quantities (\ref{loops12}) and
(\ref{loops12E}), obtained {\it after} the removal of the IR cutoff
($T \to \infty$), are still connected by the usual analytic continuation in
the angular variables only ($\chi \to i\theta$).
This is a highly non--trivial result and will be discussed in a more general
context, together with Eq. (\ref{analytic}), in Sect. 4.

\newsection{The case of QCD at order ${\cal O}(g^4)$}

\noindent
In this section we shall generalize the results obtained in the previous
section to the case of a non--Abelian gauge theory with $N_c$ colours,
but stopping to the order ${\cal O}(g^4)$ in perturbation theory.
In order to do this, let us consider the following quantity:
\ba
{\cal C}_M(\chi;~\vec{z}_\perp,\vec{R}_{1\perp},\vec{R}_{2\perp}) &\equiv&
\displaystyle\lim_{T \to \infty} \left[
{\cal G}_M(\chi;~T;~\vec{z}_\perp,\vec{R}_{1\perp},\vec{R}_{2\perp})
- 1 \right] \nonumber \\
&=& \displaystyle\lim_{T \to \infty}
{ \langle {\cal W}^{(T)}_1 (\vec{z}_\perp,\vec{R}_{1\perp})
{\cal W}^{(T)}_2 (\vec{0}_\perp,\vec{R}_{2\perp}) \rangle^{(c)} \over
\langle {\cal W}^{(T)}_1 (\vec{z}_\perp,\vec{R}_{1\perp}) \rangle
\langle {\cal W}^{(T)}_2 (\vec{0}_\perp,\vec{R}_{2\perp}) \rangle } ,
\label{C12}
\ea
where the function ${\cal G}_M$ has been defined in Eq. (\ref{GM-GE}) and
\be
\langle {\cal W}_1 {\cal W}_2 \rangle^{(c)} \equiv
\langle {\cal W}_1 {\cal W}_2 \rangle -
\langle {\cal W}_1 \rangle \langle {\cal W}_2 \rangle
\label{conn}
\ee
is defined as the {\it connected} part of the correlator $\langle {\cal W}_1
{\cal W}_2 \rangle$.
In the previous section we have evaluated the quantity (\ref{C12}) in the
{\it quenched} theory, at all orders in perturbation theory [see Eq.
(\ref{loops12})]. By expanding the result in powers of the electric charge
up to the order ${\cal O}(e^4)$, we find that:
\ba
\lefteqn{
{\cal C}_M(\chi;~\vec{z}_\perp,\vec{R}_{1\perp},\vec{R}_{2\perp}) =
\exp \left[ -i 4e^2 \coth \chi~
t(\vec{z}_\perp,\vec{R}_{1\perp},\vec{R}_{2\perp}) \right] - 1 }
\nonumber \\
& & = -i 4e^2 \coth \chi~
t(\vec{z}_\perp,\vec{R}_{1\perp},\vec{R}_{2\perp})
- 8e^4 \coth^2 \chi~
[t(\vec{z}_\perp,\vec{R}_{1\perp},\vec{R}_{2\perp})]^2 + \ldots ,
\label{pert}
\ea
where
\ba
t(\vec{z}_\perp,\vec{R}_{1\perp},\vec{R}_{2\perp}) &\equiv&
\displaystyle\int {d^2 \vec{k}_\perp \over (2\pi)^2}
{e^{-i\vec{k}_\perp \cdot \vec{z}_\perp} \over \vec{k}_\perp^2}
\sin \left( {\vec{k}_\perp \cdot \vec{R}_{1\perp} \over 2} \right)
\sin \left( {\vec{k}_\perp \cdot \vec{R}_{2\perp} \over 2} \right)
\nonumber \\
&=& {1 \over 8\pi} \ln \left(
{ |\vec{z}_\perp+{\vec{R}_{1\perp} \over 2}+{\vec{R}_{2\perp} \over 2}|
  |\vec{z}_\perp-{\vec{R}_{1\perp} \over 2}-{\vec{R}_{2\perp} \over 2}| \over
  |\vec{z}_\perp+{\vec{R}_{1\perp} \over 2}-{\vec{R}_{2\perp} \over 2}|
  |\vec{z}_\perp-{\vec{R}_{1\perp} \over 2}+{\vec{R}_{2\perp} \over 2}| }
\right) .
\label{t-function}
\ea
Of course, the same result (\ref{pert}) can also be obtained by a direct
perturbative calculation up to the order ${\cal O}(e^4)$.
This means that one first evaluates the contribution (that we shall call
${\cal M}({\rm a})$, ${\cal M}({\rm b})$, etc.) to the connected correlator
$\langle {\cal W}_1 {\cal W}_2 \rangle^{(c)}$, defined in (\ref{conn}),
coming from each diagram reported in Fig. 1; then, to obtain the quantity
(\ref{C12}), one must divide the sum of these contributions,
${\cal M}({\rm a}) + {\cal M}({\rm b}) + \ldots$, by the product of the
expectation values $\langle {\cal W}_1 \rangle$ and $\langle
{\cal W}_2 \rangle$, expanded in powers of the electric charge $e$:
\ba
\langle {\cal W}^{(T)}_1 (\vec{z}_\perp,\vec{R}_{1\perp}) \rangle &=&
1 + \overline{\cal Z}^{(2)}_{M1}~ e^2 + \ldots , \nonumber \\
\langle {\cal W}^{(T)}_2 (\vec{0}_\perp,\vec{R}_{2\perp}) \rangle &=&
1 + \overline{\cal Z}^{(2)}_{M2}~ e^2 + \ldots .
\label{Z2}
\ea
An explicit calculation shows that:
\ba
{\cal M}({\rm d}) + {\cal M}({\rm e}) + {\cal M}({\rm f}) &=&
{\cal M}({\rm a})~ \overline{\cal Z}^{(2)}_{M1}~ e^2 , \nonumber \\
{\cal M}({\rm g}) + {\cal M}({\rm h}) + {\cal M}({\rm i}) &=&
{\cal M}({\rm a})~ \overline{\cal Z}^{(2)}_{M2}~ e^2 .
\label{Md-i}
\ea
After inserting the results (\ref{Md-i}) and the expansions (\ref{Z2})
into Eq. (\ref{C12}), we conclude that:
\be
{\cal C}_M(\chi;~\vec{z}_\perp,\vec{R}_{1\perp},\vec{R}_{2\perp}) =
{\cal M}({\rm a}) + {\cal M}({\rm b}) + {\cal M}({\rm c}) + {\cal O}(e^6) ,
\label{pert2}
\ee
where ${\cal M}({\rm a})$ is the contribution [of order ${\cal O}(e^2)$] coming
from the one--photon exchange, reported in Fig. 1a, while ${\cal M}({\rm b})$
and ${\cal M}({\rm c})$ are the contributions [of order ${\cal O}(e^4)$]
coming from the two--photon exchange, reported in Figs. 1b (conventionally
called {\it ladder} diagram) and 1c (conventionally called {\it crossed}
diagram). A comparison with Eq. (\ref{pert}) gives immediately:
\ba
{\cal M}({\rm a})  &=& -i 4e^2 \coth \chi~
t(\vec{z}_\perp,\vec{R}_{1\perp},\vec{R}_{2\perp}) ,
\nonumber \\
{\cal M}({\rm b}) + {\cal M}({\rm c}) &=& 
- 8e^4 \coth^2 \chi~ [t(\vec{z}_\perp,\vec{R}_{1\perp},\vec{R}_{2\perp})]^2 .
\label{Mabc}
\ea
Now let us consider the quantity (\ref{C12}) in a non--Abelian gauge theory
with colour group $SU(N_c)$.
The perturbative diagrams to be considered for the connected correlator
(\ref{conn}) up to the order ${\cal O}(g^4)$
are the same diagrams reported in Fig. 1 for the Abelian theory, {\it plus}
other diagrams, not reported in Fig. 1, having only one gluon attachment in at
least one of the two Wilson loops, i.e., obtained by expanding at least one
of the two Wilson loops only up to the first order in $g~ A$.
All of these diagrams, with the exception of ${\cal M}({\rm b})$ and
${\cal M}({\rm c})$, vanish, by virtue of the fact that $\Tr [ T_a ] = 0$.
Viceversa, the contributions from the two--gluon exchange diagrams
${\cal M}({\rm b})$ and ${\cal M}({\rm c})$ are exactly equal to the
corresponding contributions in the Abelian theory, provided one substitutes
the electric charge $e$ with the strong coupling constant $g$ and one
multiplies by the following colour factor (the {\it same} for the two
diagrams!):
\be
{1 \over N_c} \Tr [T_a T_b] {1 \over N_c} \Tr [T_a T_b] =
{1 \over N_c} \Tr [T_a T_b] {1 \over N_c} \Tr [T_b T_a] =
{N_c^2 - 1 \over 4 N_c^2} .
\label{factor1}
\ee
Therefore, evaluating the quantity (\ref{C12}) up to the order
${\cal O}(g^4)$, one finds that:
\be
{\cal C}_M(\chi;~\vec{z}_\perp,\vec{R}_{1\perp},\vec{R}_{2\perp}) =
- 2g^4 \left( {N_c^2 - 1 \over N_c^2} \right) \coth^2 \chi~
[t(\vec{z}_\perp,\vec{R}_{1\perp},\vec{R}_{2\perp})]^2 + {\cal O}(g^6) .
\label{QCD-pert}
\ee
A completely analogous calculation in the Euclidean theory furnishes
the following result:
\be
{\cal C}_E(\theta;~\vec{z}_\perp,\vec{R}_{1\perp},\vec{R}_{2\perp}) =
2g^4 \left( {N_c^2 - 1 \over N_c^2} \right) \cot^2 \theta~
[t(\vec{z}_\perp,\vec{R}_{1\perp},\vec{R}_{2\perp})]^2 + {\cal O}(g^6) .
\label{QCD-pertE}
\ee
The results (\ref{QCD-pert}) and (\ref{QCD-pertE}) were also derived
in Refs. \cite{LLCM2,BB} using different approaches.
(See also Ref. \cite{KL} and references therein.)\\
As in the Abelian case, the Minkowskian quantity (\ref{QCD-pert}) and the
corresponding Euclidean quantity (\ref{QCD-pertE}) are connected by the
usual analytic continuation in the angular variables ($\chi \to i\theta$).
It is also easy to see, of course, that the ratios (\ref{GM-GE}), regularized
with a {\it finite} IR cutoff $T$, are connected by the analytic
continuation (\ref{analytic}) in the angular variables and in the IR
cutoff \cite{Meggiolaro02}: the proof still relies on the corresponding
results already found for the Abelian case.

\newsection{Analytic continuation and IR finitness}

\noindent
As already stated in Ref. \cite{Meggiolaro02} and in the Introduction, the
analytic continuation (\ref{analytic}) is expected to be an {\it exact}
result, i.e., not restricted to some order in perturbation theory or to some
other approximation (such as the {\it quenched} approximation), valid both
for the Abelian and the non--Abelian case.
This can be proved simply by repeating step by step the proof derived in Ref.
\cite{Meggiolaro02} for the case of Wilson lines, after adapting all the
definitions from the case of Wilson lines to the case of Wilson loops.

Let us consider, therefore, the following quantities, defined in
Minkowski space--time:
\ba
{\cal G}_M (p_1, p_2;~T;~\vec{z}_\perp,\vec{R}_{1\perp},\vec{R}_{2\perp}) &=&
{{\cal M} (p_1, p_2;~T;~\vec{z}_\perp,\vec{R}_{1\perp},\vec{R}_{2\perp})
\over
{\cal Z}_M(p_1;~T;~\vec{R}_{1\perp}) {\cal Z}_M(p_2;~T;~\vec{R}_{2\perp})} ,
\nonumber \\
{\cal M} (p_1, p_2;~T;~\vec{z}_\perp,\vec{R}_{1\perp},\vec{R}_{2\perp}) &=&
\langle {\cal W}^{(T)}_{p_1} (\vec{z}_\perp,\vec{R}_{1\perp})
{\cal W}^{(T)}_{p_2} (\vec{0}_\perp,\vec{R}_{2\perp}) \rangle ,
\nonumber \\
{\cal Z}_M(p;~T;~\vec{R}_\perp) &=&
\langle {\cal W}^{(T)}_p (\vec{z}_\perp,\vec{R}_\perp) \rangle =
\langle {\cal W}^{(T)}_p (\vec{0}_\perp,\vec{R}_\perp) \rangle .
\label{M}
\ea
The Minkowskian four--momenta $p_1$ and $p_2$ are arbitrary longitudinal
four--vectors [i.e., $\vec{p}_{1\perp} = \vec{p}_{2\perp} = \vec{0}_\perp$]
and define the longitudinal trajectories of the two IR--regularized Wilson
loops (\ref{QCDloops}), along the rectangular paths ${\cal C}_1$ and
${\cal C}_2$ defined in Eq. (\ref{traj}).

In an analogous way, we can consider the following quantities, defined
in Euclidean four--space:
\ba
{\cal G}_E (p_{1E}, p_{2E};~T;~\vec{z}_\perp,\vec{R}_{1\perp},\vec{R}_{2\perp})
&=& {{\cal E} (p_{1E}, p_{2E};~T;
~\vec{z}_\perp,\vec{R}_{1\perp},\vec{R}_{2\perp}) \over
{\cal Z}_E(p_{1E};~T;~\vec{R}_{1\perp})
{\cal Z}_E(p_{2E};~T;~\vec{R}_{2\perp})} ,
\nonumber \\
{\cal E} (p_{1E}, p_{2E};~T;~\vec{z}_\perp,\vec{R}_{1\perp},\vec{R}_{2\perp})
&=& \langle \tilde{\cal W}^{(T)}_{p_{1E}} (\vec{z}_\perp,\vec{R}_{1\perp})
\tilde{\cal W}^{(T)}_{p_{2E}} (\vec{0}_\perp,\vec{R}_{2\perp}) \rangle_E ,
\nonumber \\
{\cal Z}_E(p_E;~T;~\vec{R}_\perp) &=&
\langle \tilde{\cal W}^{(T)}_{p_E} (\vec{z}_\perp,\vec{R}_\perp) \rangle_E =
\langle \tilde{\cal W}^{(T)}_{p_E} (\vec{0}_\perp,\vec{R}_\perp) \rangle_E .
\label{E}
\ea
The Euclidean four--momenta $p_{1E}$ and $p_{2E}$ are arbitrary longitudinal
four--vectors [i.e., $\vec{p}_{1E\perp} = \vec{p}_{2E\perp} = \vec{0}_\perp$]
and define the longitudinal trajectories of the two IR--regularized Euclidean
Wilson loops, along the rectangular paths $\tilde{\cal C}_1$ and
$\tilde{\cal C}_2$ defined in Eq. (\ref{trajE}).

We can now proceed as in Ref. \cite{Meggiolaro02} and use the definition of
the path--ordered exponential in Eq. (\ref{QCDloops}) to explicitly write the
Wilson loops ${\cal W}^{(T)}_{p_1}$ and ${\cal W}^{(T)}_{p_2}$ as power series
in the exponents $g~ A$. Therefore, the quantity
${\cal M} (p_1, p_2;~T;~\vec{z}_\perp,\vec{R}_{1\perp},\vec{R}_{2\perp})$ is
defined to be the series ${\cal M} = \sum_{n=0}^\infty \sum_{r=0}^\infty
{\cal M}_{(n,r)}$, where ${\cal M}_{(n,r)}$ is the contribution from the
piece with $(g~ A)^n$ in the expansion of ${\cal W}^{(T)}_{p_1}$ and
from the piece with $(g~ A)^r$ in the expansion of ${\cal W}^{(T)}_{p_2}$;
it is given by [${\cal M}_{(0,0)} = 1$]:
\ba
\lefteqn{
{\cal M}_{(n,r)} (p_1, p_2;~T;~\vec{z}_\perp,\vec{R}_{1\perp},
\vec{R}_{2\perp}) }
\nonumber \\
& & = {(-ig)^{(n+r)} \over n! r! N_c^2}
\displaystyle\oint_{{\cal C}_1} dx_1^{\mu_1} \ldots 
\displaystyle\oint_{{\cal C}_1} dx_n^{\mu_n}
\displaystyle\oint_{{\cal C}_2} dy_1^{\nu_1} \ldots 
\displaystyle\oint_{{\cal C}_2} dy_r^{\nu_r}
\nonumber \\
& & \times \langle \Tr \left\{ {\cal P}_1 \left[ A_{\mu_n} (x_n) \ldots
A_{\mu_1} (x_1) \right] \right\}
\Tr \left\{ {\cal P}_2 \left[ A_{\nu_r} (y_r) \ldots
A_{\nu_1} (y_1) \right] \right\} \rangle ,
\label{Mnr}
\ea
where ${\cal P}_1 [\ldots]$ is the path ordering along the path ${\cal C}_1$,
while ${\cal P}_2 [\ldots]$ is the path ordering along the path ${\cal C}_2$.
Analogously, the Euclidean quantity
${\cal E} (p_{1E}, p_{2E};~T;~\vec{z}_\perp,\vec{R}_{1\perp},\vec{R}_{2\perp})$
is defined to be the series ${\cal E} = \sum_{n=0}^\infty \sum_{r=0}^\infty
{\cal E}_{(n,r)}$, where ${\cal E}_{(n,r)}$ is given by
[${\cal E}_{(0,0)} = 1$]:
\ba
\lefteqn{
{\cal E}_{(n,r)} (p_{1E}, p_{2E};~T;~\vec{z}_\perp,\vec{R}_{1\perp},
\vec{R}_{2\perp}) }
\nonumber \\
& & = {(-ig)^{(n+r)} \over n! r! N_c^2}
\displaystyle\oint_{\tilde{\cal C}_1} dx_{1E}^{\mu_1} \ldots 
\displaystyle\oint_{\tilde{\cal C}_1} dx_{nE}^{\mu_n}
\displaystyle\oint_{\tilde{\cal C}_2} dy_{1E}^{\nu_1} \ldots 
\displaystyle\oint_{\tilde{\cal C}_2} dy_{rE}^{\nu_r}
\nonumber \\
& & \times \langle \Tr \left\{ {\cal P}_1 \left[ A^{(E)}_{\mu_n} (x_{nE})
\ldots A^{(E)}_{\mu_1} (x_{1E}) \right] \right\}
\Tr \left\{ {\cal P}_2 \left[ A^{(E)}_{\nu_r} (y_{rE}) \ldots
A^{(E)}_{\nu_1} (y_{1E}) \right] \right\} \rangle_E .
\label{Enr}
\ea
Similarly, we can write the Wilson--loop expectation value
${\cal Z}_M(p;~T;~\vec{R}_\perp)$ as the series
${\cal Z}_M = \sum_{n=0}^{\infty} {\cal Z}_M^{(n)}$, where ${\cal Z}_M^{(n)}$
is the contribution from the piece with $(g~ A)^n$ in the expansion of the
Wilson loop ${\cal W}^{(T)}_{p}$; it is given by [${\cal Z}_M^{(0)} = 1$]:
\be
{\cal Z}_M^{(n)}(p;~T;~\vec{R}_\perp)
= {(-ig)^n \over n! N_c}
\displaystyle\oint_{\cal C} dx_1^{\mu_1} \ldots 
\displaystyle\oint_{\cal C} dx_n^{\mu_n}
\langle \Tr \left\{ {\cal P} \left[ A_{\mu_n} (x_n) \ldots
A_{\mu_1} (x_1) \right] \right\} \rangle .
\label{ZMn}
\ee
In the Euclidean theory we can write, analogously,
${\cal Z}_E = \sum_{n=0}^{\infty} {\cal Z}_E^{(n)}$, with
[${\cal Z}_E^{(0)} = 1$]:
\be
{\cal Z}_E^{(n)}(p_E;~T;~\vec{R}_\perp)
= {(-ig)^n \over n! N_c}
\displaystyle\oint_{\tilde{\cal C}} dx_{1E}^{\mu_1} \ldots 
\displaystyle\oint_{\tilde{\cal C}} dx_{nE}^{\mu_n}
\langle \Tr \left\{ {\cal P} \left[ A^{(E)}_{\mu_n} (x_{nE})
\ldots A^{(E)}_{\mu_1} (x_{1E}) \right] \right\} \rangle_E .
\label{ZEn}
\ee
In the expressions (\ref{Mnr}), (\ref{Enr}), (\ref{ZMn}) and (\ref{ZEn}) there
are contributions from the {\it longitudinal} segments of the paths, which can
be treated in the same way adopted in Ref. \cite{Meggiolaro02} for the case of
Wilson lines; but there are also contributions from the {\it transverse}
segments (connecting the longitudinal segments) of the paths, which however,
as we shall see below, do not alter the conclusions found in Ref.
\cite{Meggiolaro02}.
Let us consider, for example, a contribution from the transverse segment at
$\tau = +T$ in ${\cal C}_1$, in the Minkowskian case:
\be
\ldots (-ig) \displaystyle\int_{-1}^{+1} dv \left( -{R_1^\mu \over 2} \right)
\ldots \langle \ldots A_\mu ( z + T{p_1 \over m}
- v{R_1 \over 2} ) \ldots \rangle ,
\label{transM}
\ee
and the corresponding contribution from the transverse segment at $\tau = +T$
in $\tilde{\cal C}_1$, in the Euclidean case:
\be
\ldots (-ig) \displaystyle\int_{-1}^{+1} dv \left( -{R_{1E\mu} \over 2} \right)
\ldots \langle \ldots A^{(E)}_\mu ( z_E + T{p_{1E} \over m}
- v{R_{1E} \over 2} ) \ldots \rangle_E .
\label{transE}
\ee
Making use of the correspondence between the Minkowskian and the Euclidean
gluonic Green functions, i.e.,
\be
\ldots B_{E\mu} \ldots \langle \ldots A_\mu^{(E)} (x_E) \ldots \rangle_E =
\ldots \tilde{B}^\mu \ldots \langle \ldots A_\mu (\tilde{x}) \ldots \rangle ,
\label{Green0}
\ee
where $x_E = (\vec{x}_E, x_{E4})$ are Euclidean four--coordinates and
$B_E = (\vec{B}_E, B_{E4})$ is any Euclidean four--vector, while
$\tilde{x}$ and $\tilde{B}$ are Minkowskian four--vectors defined as
\be
\tilde{x} = (\tilde{x}_0, \vec{\tilde{x}}) \equiv
(-i x_{E4}, \vec{x}_E) ,~~~~
\tilde{B} = (\tilde{B}^0, \vec{\tilde{B}}) \equiv
(-i B_{E4}, \vec{B}_E) ,
\label{4vect}
\ee
in our specific case we can state that:
\ba
\ldots (-ig) \displaystyle\int_{-1}^{+1} dv \left( -{R_{1E\mu} \over 2} \right)
\ldots \langle \ldots A^{(E)}_\mu ( z_E + T{p_{1E} \over m}
- v{R_{1E} \over 2} ) \ldots \rangle_E
\nonumber \\
= \ldots (-ig) \displaystyle\int_{-1}^{+1} dv \left( -{R_1^\mu \over 2} \right)
\ldots \langle \ldots A_\mu ( z -iT{\bar{p}_1 \over m}
- v{R_1 \over 2} ) \ldots \rangle ,
\label{Green1}
\ea
where $p_{kE} = (\vec{p}_{kE}, p_{kE4})$, for $k = 1,2$, are the two Euclidean
four--vectors introduced above while $\bar{p}_k$ are two corresponding
Minkowskian four--vectors, defined as:
\be
\bar{p}_k \equiv i\tilde{p}_k = (p_{kE4}, i\vec{p}_{kE}) ,
~~~ {\rm for} ~k=1,2 .
\label{pbar}
\ee
[In deriving Eq. (\ref{Green1}) we have used the fact that, using Eq.
(\ref{4vect}), $\tilde{z} = z$, being $z = (0,0,\vec{z}_\perp)$ and
$z_E = (0,\vec{z}_\perp,0)$, and, similarly, $\tilde{R}_1 = R_1$ and
$\tilde{R}_2 = R_2$.]

Eq. (\ref{Green1}), when compared with Eq. (\ref{transM}), implies that the
same type of analytic continuation, which was found in Ref. \cite{Meggiolaro02}
considering the case of (longitudinal) Wilson lines, also holds between the
Euclidean and the Minkowskian amplitudes constructed with Wilson loops; i.e.,
\be
{\cal E} (p_{1E}, p_{2E};~T;~\vec{z}_\perp,\vec{R}_{1\perp},\vec{R}_{2\perp}) =
{\cal M} (\bar{p}_1, \bar{p}_2;~-iT;~\vec{z}_\perp,\vec{R}_{1\perp},
\vec{R}_{2\perp}) ,
\label{EMres2}
\ee
and also, for the Wilson--loop expectation values:
\be
{\cal Z}_E (p_E;~T;~\vec{R}_\perp) = {\cal Z}_M (\bar{p};~-iT;~\vec{R}_\perp) ,
\label{ZEMres2}
\ee
where, as usual, $p_E = (\vec{p}_E, p_{E4})$
and $\bar{p} \equiv i\tilde{p} = (p_{E4}, i\vec{p}_E)$.

Finally, in the Minkowskian theory, we {\it choose} $p_1$ and $p_2$ to be the
four--momenta (\ref{p1p2}) of two particles with mass $m$, moving with speed
$\beta$ and $-\beta$ along the $x^1$--direction.
Analogously, in the Euclidean theory we {\it choose}
a reference frame in which the spatial vectors $\vec{p}_{1E}$ and 
$\vec{p}_{2E} = -\vec{p}_{1E}$ are along the $x_{E1}$--direction and,
moreover, $p_{1E}^2 = p_{2E}^2 = m^2$: $p_{1E}$ and $p_{2E}$ are given by
Eq. (\ref{p1p2E}).
With this choice, one has that:
\ba
\bar{p}_1 &=&
m (\cos {\theta \over 2},i\sin {\theta \over 2},0,0) =
m (\cosh {i\theta \over 2},\sinh {i\theta \over 2},0,0) , \nonumber \\
\bar{p}_2 &=&
m (\cos {\theta \over 2},-i\sin {\theta \over 2},0,0) =
m (\cosh {i\theta \over 2},-\sinh {i\theta \over 2},0,0) .
\label{pbar12}
\ea
A comparison with the expressions (\ref{p1p2}) for the Minkowskian
four--vectors $p_1$ and $p_2$ reveals that $\bar{p}_1$ and $\bar{p}_2$ are
obtained from $p_1$ and $p_2$ after the analytic continuation
$\chi \rightarrow i \theta$ in the angular variables is made.
Therefore, if we denote with ${\cal M}(\chi;~T;~\vec{z}_\perp,\vec{R}_{1\perp},
\vec{R}_{2\perp})$ the value of ${\cal M}(p_1, p_2;~T;~\vec{z}_\perp,
\vec{R}_{1\perp},\vec{R}_{2\perp})$ for $p_1$ and $p_2$ given by Eq.
(\ref{p1p2}) and we also denote with ${\cal E}(\theta;~T;~\vec{z}_\perp,
\vec{R}_{1\perp},\vec{R}_{2\perp})$ the value of
${\cal E}(p_{1E}, p_{2E};~T;~\vec{z}_\perp,\vec{R}_{1\perp},\vec{R}_{2\perp})$
for $p_{1E}$ and $p_{2E}$ given by Eq. (\ref{p1p2E}), we find, using the result
(\ref{EMres2}) derived above:
\be
{\cal E}(\theta;~T;~\vec{z}_\perp, \vec{R}_{1\perp},\vec{R}_{2\perp}) =
{\cal M}(\chi \to i\theta;~T \to -iT;~\vec{z}_\perp,\vec{R}_{1\perp},
\vec{R}_{2\perp}) .
\label{EMres4}
\ee
Let us consider, now, the Wilson--loop expectation value, i.e.,
${\cal Z}_M(p;~T;~\vec{R}_\perp)$ in the Minkowskian theory and
${\cal Z}_E(p_E;~T;~\vec{R}_\perp)$ in the Euclidean theory.
Clearly, ${\cal Z}_M(p;~T;~\vec{R}_\perp)$, considered as a function of a
general (longitudinal) four--vector $p$, is a scalar function constructed 
with the only four--vector $u = p/m$ (being $p \cdot R = 0$).
Similarly, ${\cal Z}_E(p_E;~T;~\vec{R}_\perp)$, considered as a function of a
general (longitudinal) Euclidean four--vector $p_E$, is a scalar function
constructed with the only four--vector $u_E = p_E/m$
(being $p_E \cdot R_E = 0$).
Therefore, if we denote with  ${\cal Z}_M(T;~\vec{R}_\perp)$ the value
of ${\cal Z}_M(p_1;~T;~\vec{R}_\perp)$ or ${\cal Z}_M(p_2;~T;~\vec{R}_\perp)$,
for $p_1$ and $p_2$ given by Eq. (\ref{p1p2}) [being $p_1^2 = p_2^2 = m^2$],
and we also denote with ${\cal Z}_E(T;~\vec{R}_\perp)$ the value
of ${\cal Z}_E(p_{1E};~T;~\vec{R}_\perp)$ or
${\cal Z}_E(p_{2E};~T;~\vec{R}_\perp)$ for $p_{1E}$ and $p_{2E}$ given by Eq.
(\ref{p1p2E}) [being: $p_{1E}^2 = p_{2E}^2 = m^2$], we find, using the result
(\ref{ZEMres2}) derived above (and the fact that $\bar{p}_1^2 = \bar{p}_2^2
= m^2$):
\be
{\cal Z}_E(T;~\vec{R}_\perp) = {\cal Z}_M(-iT;~\vec{R}_\perp) .
\label{ZEMres4}
\ee
Combining this identity with Eq. (\ref{EMres4}), we find that the Minkowskian
and the Euclidean amplitudes, defined by Eqs. (\ref{M}) and (\ref{E}),
with $p_1$ and $p_2$ given by Eq. (\ref{p1p2}) and $p_{1E}$ and $p_{2E}$
given by Eq. (\ref{p1p2E}),
are connected by the following analytic continuation in the angular variables
and in the IR cutoff:
\ba
{\cal G}_E(\theta;~T;~\vec{z}_\perp,\vec{R}_{1\perp},\vec{R}_{2\perp}) &=&
{\cal G}_M(\chi \to i\theta;~T \to -iT;
~\vec{z}_\perp,\vec{R}_{1\perp},\vec{R}_{2\perp}) ,
\nonumber \\
{\rm i.e.,}~~
{\cal G}_M(\chi;~T;~\vec{z}_\perp,\vec{R}_{1\perp},\vec{R}_{2\perp}) &=&
{\cal G}_E(\theta \to -i\chi;~T \to iT;
~\vec{z}_\perp,\vec{R}_{1\perp},\vec{R}_{2\perp}) .
\label{final}
\ea
We have derived the relation (\ref{final}) of analytic continuation for a
non--Abelian gauge theory with gauge group $SU(N_c)$. It is clear, from the
derivation given above, that the same result is valid also for an Abelian
gauge theory (QED).

In the previous sections we have proved [in the case of {\it quenched} QED
and in the case of QCD at order ${\cal O}(g^4)$] that the loop--loop
correlation functions introduced above, both in the Minkowskian and in the
Euclidean theory, are indeed IR--finite quantities, i.e., they have finite
limits when letting $T \to \infty$.
This IR finitness of the loop--loop correlators was first argued in Ref.
\cite{JP} simply on the basis of (sensible) physical arguments and has been
found to be correct in all different approaches for the calculation of the
correlators (even if not in a direct way, but using various conjectures or
models) appeared up to now in the literature (we shall briefly recall some
of them in the next section).
One can then define the loop--loop correlation function (\ref{C12}) in the
Minkowskian theory and the corresponding quantity in the Euclidean theory,
with the IR cutoff removed,
and one finds that they are connected by the usual analytic continuation
in the angular variables:
\ba
{\cal C}_E(\theta;~\vec{z}_\perp,\vec{R}_{1\perp},\vec{R}_{2\perp}) &=&
{\cal C}_M(\chi \to i\theta;~\vec{z}_\perp,\vec{R}_{1\perp},\vec{R}_{2\perp}) ,
\nonumber \\
{\rm i.e.,}~~
{\cal C}_M(\chi;~\vec{z}_\perp,\vec{R}_{1\perp},\vec{R}_{2\perp}) &=&
{\cal C}_E(\theta \to -i\chi;~\vec{z}_\perp,\vec{R}_{1\perp},\vec{R}_{2\perp}) .
\label{final-C}
\ea
Clearly, if ${\cal G}_M$ and ${\cal G}_E$ are analytic functions of $T$ in the
whole complex plane and if $T=\infty$ is an ``eliminable singular point''
[i.e., the finite limit (\ref{C12}) exists when letting the {\it complex}
variable $T \to \infty$], then, of course, the analytic continuation
(\ref{final-C}) immediately derives from Eq. (\ref{final}), when letting
$T \to +\infty$. (For example, if ${\cal G}_M$ and ${\cal G}_E$ are analytic
functions of $T$ and they are bounded at large $T$, i.e., $\forall R_{M,E}
\in \Re^+,~\exists B_{M,E} \in \Re^+$ such that
$|{\cal G}_{M,E}(T)| < B_{M,E}$ for $R_{M,E} < |T| < \infty$, then $T=\infty$
is an ``eliminable singular point'' for both of them.)
But the same result (\ref{final-C}) can also be derived under weaker
conditions. For example, let us assume that ${\cal G}_E$ is a bounded
analytic function of $T$ in the sector $0 \le \arg T \le {\pi \over 2}$,
with finite limits along the two straight lines on the border of the sector:
${\cal G}_E \to G_{E1}$, for $({\rm Re}T \to +\infty,~{\rm Im}T = 0)$, and
${\cal G}_E \to G_{E2}$, for $({\rm Re}T = 0,~{\rm Im}T \to +\infty)$.
And, similarly, let us assume that ${\cal G}_M$ is a bounded
analytic function of $T$ in the sector $-{\pi \over 2} \le \arg T \le 0$,
with finite limits along the two straight lines on the border of the sector:
${\cal G}_M \to G_{M1}$, for $({\rm Re}T \to +\infty,~{\rm Im}T = 0)$, and
${\cal G}_M \to G_{M2}$, for $({\rm Re}T = 0,~{\rm Im}T \to -\infty)$.
We can then apply the ``Phragm\'en--Lindel\"of theorem'' (see, e.g., the
theorem 5.64 in Ref. \cite{PLT}) to state that $G_{E2} = G_{E1}$ and
$G_{M2} = G_{M1}$. Therefore, also in this case, the analytic continuation
(\ref{final-C}) immediately derives from Eq. (\ref{final}) when $T \to \infty$.

\newsection{Concluding remarks and prospects}

\noindent
Eq. (\ref{final-C}) is the main result of this paper and, as we have argued
in the previous sections, its general validity is based: i) on the relation
(\ref{final}) of analytic continuation in the angular variables and in the
IR cutoff, which we have proved to be an exact and general result,
and ii) on the IR--finitness of the loop--loop correlation functions,
which we have explicitly tested in some special cases in Sects. 2 and 3 and,
as we have said at the end of the previous section, is expected to be a
general result too. (Indeed, the validity of the relation (\ref{final-C}) has
been also recently verified in Ref. \cite{BB} by an explicit calculation
up to the order ${\cal O}(g^6)$ in perturbation theory.)

The relation (\ref{final-C})
allows the derivation of the {\it loop--loop scattering amplitude}
(\ref{scatt-loop}) from the analytic continuation $\theta \to -i\chi$ of the
corresponding Euclidean quantity, which can be evaluated non perturbatively by
well--known and well--established techniques available in the Euclidean theory.
Finally, in order to obtain the correct $s$--dependence of the scattering
amplitude, one must express the hyperbolic angle $\chi$ between the two loops
in terms of $s$, in the high--energy limit $s \to \infty$, i.e.,
$\chi \sim \log \left( {s \over m_1 m_2} \right)$,
where $m_1$ and $m_2$ are the masses of the two hadrons considered.\\
This approach has been extensively used in the literature in order to
address, from a theoretical point of view, the still unsolved problem
of the asymptotic $s$--dependence of hadron--hadron elastic scattering
amplitudes and total cross sections.\\
For example, this approach has been recently adopted
in Refs. \cite{JP,Janik}, in order to study the high--energy scattering
in strongly coupled $SU(N_c)$ gauge theories, in the large--$N_c$ limit,
using the AdS/CFT correspondence,
in Refs. \cite{instanton1,instanton2}, in order to investigate
instanton--induced effects in QCD high--energy scattering,
and also in Ref. \cite{LLCM2},
in the context of the so--called ``loop--loop correlation model'' \cite{LLCM1},
in which the QCD vacuum is described by perturbative gluon exchange
and the non--perturbative ``Stochastic Vacuum Model'' (SVM).

Both the instanton approach \cite{instanton1,instanton2} (but see also
\cite{instanton3}!) and the SVM approach \cite{LLCM2} (but see also
\cite{LLCM1}!) result in a loop--loop scattering amplitude ${\cal T}_{(ll)}$
linearly rising with $s$. By virtue of the ``optical theorem'',
this should imply (apart from possible $s$--dependences in the hadron wave
functions! See, for example, Refs. \cite{Dosch,Berger}) $s$--independent
hadron--hadron total cross sections in the
asymptotic high--energy limit, in apparent contradiction to the
experimental observations, which seem to be well described by a
``{\it pomeron}--like'' high--energy behaviour
(see, for example, Ref. \cite{pomeron-book} and references therein.
We must recall, however, that the {\it pomeron} violates the Froissart bound
\cite{FLM} and therefore must be regarded as a {\it pre--asymptotic}, but
not really {\it asymptotic}, high--energy behaviour: see, e.g., Refs.
\cite{BB,Kaidalov} and references therein).
A {\it pomeron}--like behaviour was instead found (apart from possible
undetermined $\log s$ prefactors) in the above--mentioned Refs.
\cite{JP,Janik}, using the AdS/CFT correspondence.
(It has been also recently proved in Ref. \cite{BB}, by an explicit
perturbative calculation, that the loop--loop scattering amplitude approaches,
at sufficiently high energy, the BFKL--{\it pomeron} behaviour \cite{BFKL}.)

An independent non--perturbative approach would be surely welcome and could be
provided by a direct lattice calculation of the loop--loop Euclidean
correlation functions. Clearly a lattice approach can at most furnish
only a discrete set of $\theta$--values for the above--mentioned functions,
from which it is clearly impossible (without some extra assumption on the
interpolating continuous functions) to get, by the analytic continuation
$\theta \to -i \chi$, the corresponding Minkowskian correlation functions.
However, the lattice approach could furnish a criterion to
investigate the goodness of a given existing analytic model (such as:
Instantons, SVM, AdS/CFT, BFKL and so on $\ldots$) or even to open the way
to some new model, simply by trying to fit the lattice data with the
considered model.
This would surely result in a considerable progress along this line
of research.

\vfill\eject

{\renewcommand{\Large}{\normalsize}
}

\vfill\eject

\noindent
\begin{center}
{\bf FIGURE CAPTIONS}
\end{center}
\vskip 0.5 cm
\begin{itemize}
\item [\bf Fig.~1.] The Feynman diagrams contributing in {\it quenched}
Abelian theory to the connected correlator $\langle {\cal W}_1 {\cal W}_2
\rangle^{(c)}$, defined in (\ref{conn}), up to the fourth order in the
coupling constant. The two continuum lines correspond to (portions of)
the Wilson loops ${\cal W}_1$ and ${\cal W}_2$.
The two--gluon--exchange diagrams 1b and 1c are the only ones which also
contribute in {\it full} (i.e., {\it unquenched}) non--Abelian gauge theory
up to the fourth order in the coupling constant.
\end{itemize}

\vfill\eject

\pagestyle{empty}

\centerline{\bf Figure 1}
\vskip 4truecm
\begin{figure}[htb]
\vskip 4.5truecm
\includegraphics{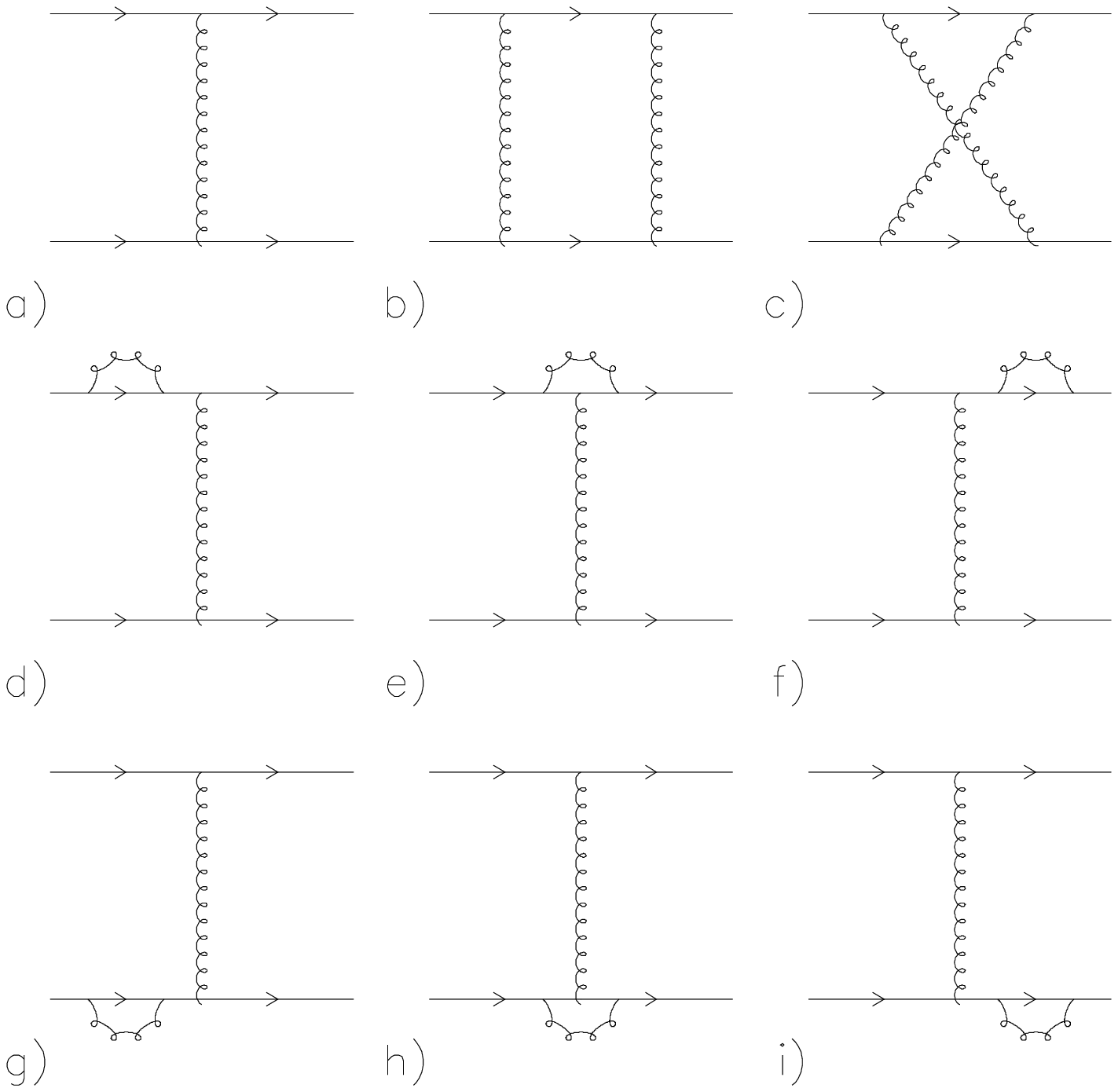}
\end{figure}

\end{document}